\documentstyle[aps,preprint,pra,epsfig]{revtex}

\begin{document}

\title{Topology Induced Spatial Bose-Einstein Condensation for Bosons on
Star-Shaped Optical Networks}

\author{I. Brunelli$^1$, G. Giusiano$^1$, F. P. Mancini$^1$, P.
Sodano$^1$, and A. Trombettoni$^2$}
\address{$^1$ Dipartimento di Fisica and Sezione I.N.F.N.,
\\Universit\`a di Perugia, Via A. Pascoli
Perugia, I-06123,  Italy}
\address{$^2$ I.N.F.M. and Dipartimento di Fisica, \\Universit\`a
di Parma, parco Area delle Scienze 7A
Parma, I-43100, Italy}

\date{\today}

\maketitle

\hyphenation{periodically pe-ri-od-i-cal-ly}

\begin{abstract}
New coherent states may be induced by pertinently engineering
the topology of a network.
As an example, we consider the properties of non-interacting
bosons on a star network, which may be realized 
with a dilute atomic gas in a star-shaped deep optical lattice. 
The ground state is localized around the star center 
and it is macroscopically occupied below the Bose-Einstein condensation 
temperature $T_c$. We show that $T_c$ depends only
on the number of the star arms and on the Josephson energy of the
bosonic Josephson junctions and that the non-condensate fraction 
is simply given by the reduced temperature $T/T_c$. 
\end{abstract}

\maketitle

\section{Introduction}
Although it is well known that free bosons hopping on translationally
invariant networks cannot undergo Bose-Einstein condensation at finite
temperature if the space dimension $d$ is less or equal to two (see
Ref. \cite{pitaevskii03}), very recent studies
\cite{burioni00,burioni01} hint to the exciting possibility  that the
network topology may act as a catalyst for inducing a finite temperature
spatial Bose-Einstein condensation even if $d<2$. As an example of this
situation we shall investigate the properties of non-interacting bosons
hopping on a star shaped optical network, evidencing that - already for
this very simple graph topology - one may have a macroscopic occupation
of the ground-state at low temperatures.

\begin{figure}[ht]
\centerline{\psfig{file=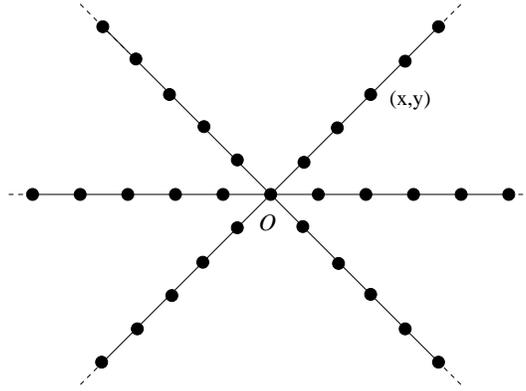,width=7cm}}
\vspace*{8pt}
\caption{A star network with six arms.}
\label{fig1}
\end{figure}

A star graph (see Fig. \ref{fig1}) is made of $p$ one-dimensional chains
({\em arms}) which merge in one point called the center of the
star. Each site $i$ of the star arms is naturally labeled by
two integer indices $x$ and $y$ where $x=0,\cdots, L$ labels the
distance from the center and $y=1,\cdots,p$  labels the arms. The center
of the network is denoted by $O \equiv (0,y)$.
The total number of sites is $N_s=(pL+1)$; each site on the arm is
linked only to two neighbors whereas the center has $p$ neighbors: thus
the fact that the center has coordination number $p$ is the source of
spatial inhomogeneity in this lattice. In the following we shall
evidence that bosons hopping on this graph undergo - at a certain
temperature $T_c$, which depends on the number of star arms - a topology
induced spatial Bose-Einstein condensation in a state localized around
the center of the star graph.

Bosons hopping on star-shaped networks
can be experimentally realized loading a dilute Bose-Einstein condensate 
(BEC) in a suitable periodic potential, arranged to provide a star-like
configuration. Periodic potentials are, nowadays, routinely created
with two or more counterpropagating laser beams; one can accurately tune
the height of the potential (which is proportional to the power
of the lasers) as well as the distance between neighboring sites.
When one loads an atomic BEC in a deep
optical lattice, one has a bosonic Josephson network,
i.e., an array of bosonic Josephson junctions (BJJ's).
A single BJJ may be obtained by loading a BEC in a double well 
potential: the weak link between the atomic condensates is provided 
by the energy barrier between them and the dynamics of the atoms
at $T=0$ is described by Josephson equations, obtained 
from the Gross-Pitaevskii equation \cite{smerzi97}.
Similarly, with a multi-well periodic potential, when
the heights of the barriers are much higher than the chemical potential, 
the system realizes a lattice of weakly coupled condensates. Each
bosonic Josephson junction consists of a pair of neighboring condensates: the
tunneling rate (proportional to the Josephson energy) is easily
tuned by changing the power of the lasers, and it decreases if the
interwell barriers increase. The properties of a bosonic Josephson 
network at $T=0$ are described by a discrete nonlinear 
Schr\"odinger equation, obtained from the Gross-Pitaevskii equation 
with a periodic potential \cite{trombettoni01}. We remark that recently 
it has been showed that a $2D$ optical network of BEC can be 
described by the Gross-Pitaevskii equation also at finite temperature,
provided that the interwell energy barriers and the frequency
of the axial confinement are large enough \cite{trombettoni03}.

Present-day BJJ networks are built on
regular geometries: linear chains \cite{anderson98}, 
squares \cite{greiner01} and cubes \cite{greiner02}. However, 
a variety of non-conventional structures may be produced
by the standing waves of several interfering laser beams suitably placed 
\cite{oberthaler02}. For instance, a star with four arms may be realized
by having two perpendicular gaussian laser beams 
superimposed to a $2D$ optical lattice; 
the center of the star should be then arranged by adjusting
the distance between neighboring wells.

BJJ networks may be pertinently described by the 
Bose-Hubbard Hamiltonian \cite{fisher89,jaksch98}.
In fact, when all the relevant physical parameters are small
compared to the excitation energies, the field operator \cite{jaksch98},
describing the condensate configuration in the BJJ network, may be
expanded as $\hat{\psi}(\vec{r},\tau)=\sum_{j}\hat{a}_j(\tau)
\phi_j(\vec{r})$ with $\phi_j(\vec{r})$
the normalized Wannier wavefunction localized in the $j$-th well and 
$\hat{n}_j=\hat{a}^{\dag}_j\hat{a}_j$ the bosonic number operator. 
Substituting the expansion for $\hat{\psi}(\vec{r},\tau)$ in the full quantum 
Hamiltonian describing the bosonic system \cite{jaksch98} leads to a 
Bose-Hubbard model 
\begin{equation}
H=-t \sum_{<i,j>} (\hat{a}^{\dag}_i \hat{a}_j+h.c.)+\frac{U}{2}
\sum_{j}\hat{n}_j\left(\hat{n}_j-1 \right) .
\label{B-H}
\end{equation}
In Eq. (\ref{B-H}) $\sum_{<i,j>}$ denotes a sum over all the distinct
pairs of neighboring sites, $\hat{a}^{\dag}_j$ ($\hat{a}_j$) is the 
bosonic operator 
which creates (destroys) a boson at site $j$, $U=(4\pi \hbar^2a/m) \int
d\vec{r}\phi_j^4$ ($a$ is the s-wave scattering length 
and $m$ is the atomic mass) and 
$t \simeq - \int d\vec{r} \, \big[ \frac{\hbar^2}{2m} 
\vec{\nabla} \phi_i \cdot \vec{\nabla} \phi_j + 
\phi_i  V_{ext} \phi_j \big]$, where $ \phi_i$ and  $\phi_j$ are 
the Wannier functions at the neighboring sites $i$ and $j$ and 
$V_{ext}$ is the external potential confining the bosons. 
Since, for atomic condensates, $U$ may be varied
by tuning the scattering length using Feshbach resonances
\cite{pitaevskii03}, 
one may assume that $U \ll t$; for the sake of simplicity, in the
following we shall set $U=0$.

The filling, i.e., the average number of particles per site, is defined as
$f=N_T/N_s$, where $N_T$ is the total number of bosons. When $f
\gg 1$ and the fluctuations of the particle numbers per site are much
smaller than $f$, one can safely substitute the operator $\hat{a}_i$
with $\sqrt{N_i} \: e^{i \phi_i}$, where $N_i$ is the number
of particles at the site $i$ \cite{vanotterlo93,anglin01}. 
As a result, the Josephson energy of a single BJJ is given by
\begin{equation} 
E_J \approx 2 t f.
\label{Jos-ij}
\end{equation}

It is worthwhile to remark that the Bose-Hubbard Hamiltonian (\ref{B-H}) 
describes not only BEC's, but in general cold bosons 
in a deep periodic potential, provided 
that the temperature does not excite higher bands: 
$E_{gap} \gtrsim k_B T$, where $E_{gap} \propto \sqrt{V_0}$ is the 
band energy gap and $V_0$ is the interwell energy barrier.
In this paper we are considering a BEC in a deep 
optical lattice: thus we are also assuming that in each well one has 
a grain condensate, i.e., that one works at a temperature smaller than
the temperature $T_{BEC}$ at which condensation occurs in a single well.
For $T<T_{BEC}$, one has a network of weakly coupled condensates, and
the bosons, at the temperature $T_c$ defined by Eq.
(\ref{ris_temp_crit}), start to macroscopically occupy the ground-state
whose wavefunction is given by Eq. (\ref{ground_state_star}). We observe
that for $V_0 \sim 15 E_R$, where $E_R=h^2/2m \lambda^2$ with $\lambda$
the wavevector of the lattice beams, and for an average number of
particles per site $f \sim 200$, one has $T_{BEC} \sim 500 nK$ and $T_c
\sim 50 nK$ \cite{trombettoni03}. The experimental situation in which
the total number of particles is so small that $T_{BEC}$ is much
lower than $T_c$ should be also possible: in this
case for $T>T_c$ one has not an array of BJJ's, but at $T_c$ the bosons
can still macroscopically occupy the ground-state described by the
eigenfunction (\ref{ground_state_star}).

The plan of the paper is the following. In Sec. II we analyze
the spectrum of bosons hopping on a star-shaped graph; we
shall evidence that, due to the network inhomogeneity, an isolated
localized ground-state appears in the spectrum (its effect is similar
to the one induced by an impurity on a linear chain, see e.g. Ref.
\cite{economou83}); we shall also determine, as a function of the number
of star arms $p$, the energy gap between the ground-state and the
continuum states as well as the coherence length $\xi$ which
characterizes the topology induced spatial condensation of bosons in the
center of the star graph. In Sec. III we analyze the thermodynamic
properties of non-interacting bosons hopping on the star graph and we
determine the critical temperature $T_c$ as a function of $p$ and of the
Josephson energy of the single BJJ; we show that the condensate fraction
and the inhomogeneous spatial distribution of bosons over the array can
be expressed in a simple way as a function of the scaled temperature
$T/T_c$. Section IV is devoted to some remarks on the results of
our investigation and to a discussion of experimental settings which
should enable to evidence the existence of the topology induced spatial
BEC analyzed in this paper.

\section{Spectrum of bosons hopping on a star graph}

The topology and geometry of a generic graph network is fully described
by its adjacency matrix $A_{x,y;\:x',y'}$ whose entries are  $1$ if
$(x,y; \: x',y')$  is an allowed link and $0$ otherwise; for a graph
network, the Hamiltonian (\ref{B-H}) is  written as
\begin{equation}
H=-t\sum_{x,y;x',y'}
A_{x,y;\:x',y'}\: \hat{a}^{\dag}_{x,y} \hat{a}_{x',y'}.
\label{pure_hopping}
\end{equation}
The single-particle energy spectrum on a star network is found then by
solving the eigenvalue equation \cite{burioni01}:
\begin{equation}
-t \sum_{x',y'} A_{x,y;\:x',y'}\: \psi_E(x',y')= E
\psi_E(x,y) ,
\label{eigen_eq}
\end{equation}
with the adjacency matrix $A_{x,y;\:x',y'}$  given by:
\begin{equation}
A_{x,y;\:x',y'}=
(\delta_{x',x-1}+\delta_{x',x+1}) \;(1-\delta_{x,0})\; \delta_{y,y'}+
\delta_{x,0}\; \delta_{x',1} .
\label{adjacency}
\end{equation}

We shall refer the interested reader to the Appendix for the
mathematical details of the solution of the eigenvalue equation
(\ref{eigen_eq}) with adjacency matrix (\ref{adjacency}); here, we only
describe the properties of the spectrum which turn out to be relevant
for our subsequent analysis.

The spectrum $\sigma$ is formed by $N_s$ states and is divided in
three parts: $E_0$, $\sigma_{0}$ and $E_+$. $\sigma_0$ describes the
delocalized states with energies between $-2t$ and $2t$; in the
thermodynamic limit $L \to \infty$, the normalized density of these
states is given by
\begin{equation}
\rho(E)= \frac{1}{\pi \:
\sqrt{4t^2-E^2}}\:,
\label{rho_catena}
 \end{equation}
just as for a particle hopping on a linear chain.
Apart from the $(pL-1)$ continuum states belonging to $\sigma_0$, there
are two bound states confined away from the continuum corresponding
to energies $E_0 <-2t$ and $E_+>2t$. These two
eigenstates are localized and form the so-called {\em hidden} spectrum
\cite{burioni00,burioni01,buonsante02}: hidden means here that the two
states - in the thermodynamic limit - do not contribute to the
normalized density of states yielding the closure relation.

$E_0$ is the ground-state energy which, in the thermodynamic limit and
for a star graph with $p$ arms, is given by
\begin{equation}
E_0=-t\: \frac{p}{\sqrt{p-1}}.
\label{energia_ground_state}
\end{equation}
Equation (\ref{energia_ground_state}) reproduces exactly the known
result $E_0=-2t$ for $p=2$ and it implies the well known 
fact that, for a linear
chain, there are no localized states.
The energy of the isolated eigenstate in the high-energy region is
simply given by $E_+ = \mid E_0 \mid$.

\begin{figure}[ht]
\centerline{\psfig{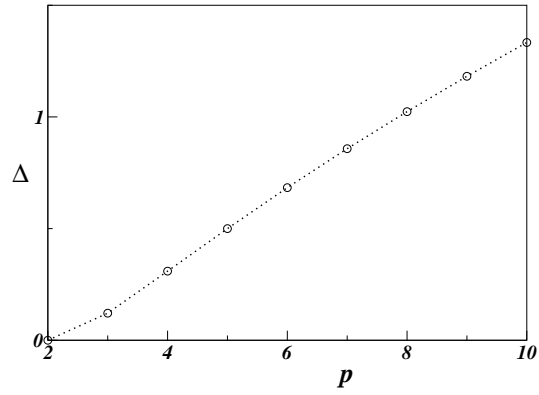}}
\vspace*{8pt}
\caption{Energy gap $\Delta$ between the ground-state energy and 
the continuum part of the spectrum (in units of the tunneling rate $t$) 
as a function of the number of arms $p$. For $p=2$ (i.e., a linear chain) 
$\Delta=0$. The dotted line is just a guide to the eye to connect the
points.}
\label{fig2}
\end{figure}

It is worth observing that the spectrum is gapped: in fact, there is a
finite gap $\Delta=\vert E_0 \vert - 2t$ between the ground-state
energy and the continuum part of the spectrum; the value of  $\Delta$
depends on the number of arms $p$. In Fig. \ref{fig2} we plot the
energy gap as a function of the number of arms. As expected, one has
$\Delta=0$ when $p=2$.

The ground-state and the eigenstate corresponding to the eigenvalue
$E_{+}$ - due to the topology of the array - are localized in the
center and exhibit an exponential decay in the direction of the
arms. In the thermodynamic limit, the normalized ground-state
wavefunction, as a function of the number of the arms
$p$,  is given by:
\begin{equation}
\label{ground_state_star}
\psi_{E_0}(x,y)=\sqrt{\frac{p-2}{2p-2}}e^{- x/\xi}
\end{equation}
(the normalization is, of course, over the whole network: i.e.,
$\sum_{x,y} \vert \psi_{E_0}(x,y) \vert^2=1$). $\xi$  provides an
estimate of the ground-state localization and it is given by
\begin{equation}
\label{corr_length}
\xi=\frac{2}{\log{(p-1)}}.
\end{equation}
In Fig. \ref{fig3} we plot the ground-state wavefunction
for different values of the number of arms $p$.
Figure \ref{fig3} evidences that adding arms enhances the
localization of the wavefunction around the center of the star graph.
As we shall see in the next section, bosons are allowed to spatially
condense in this ground-state at low temperatures.

\section{Thermodynamics of bosons hopping on a star graph}

The thermodynamical properties of non-interacting
bosons hopping on a star-graph hint to the possibility of a topology induced
spatial BEC in the center of the star graph \cite{burioni01}.
To elucidate this phenomenon, it is most convenient to introduce
the macrocanonical ensemble to determine the
fugacity $z$ as a function of
the temperature of the system. The
equation determining $z$ is given by
\begin{equation} 
N_T=\sum_{E\in \sigma}
\frac{d(E)}{z^{-1} e^{\beta (E-E_0)}-1}.
\label{filling1}
\end{equation}
In Eq. (\ref{filling1}) $d(E)$ is the degeneracy of each
single-particle eigenstate, $E_0$ is the energy of the ground-state of
the Hamiltonian (\ref{pure_hopping}) and $\beta=1/k_B T$; the sum in 
Eq. (\ref{filling1}) is over the entire spectrum $\sigma$. For free bosons
hopping on a star graph, one has
\begin{equation}
N_T=N_{E_0}+N_{E_{+}}+ \int_{E \in \sigma_0} dE
\frac{N_s \rho(E)}{z^{-1} e^{\beta (E-E_0)}-1 },
\label{filling}
\end{equation}
where $N_{E_0}(p,L;T)$ and  $N_{E_{+}}(p,L;T)$
denote, respectively,  the number of bosons which occupy the
ground-state and the state of energy $E_{+}$  at a certain temperature
$T$. $\rho(E)$, with $E\in \sigma_0$, is the density of delocalized
states defined in Eq. (\ref{rho_catena}). It is pertinent to define also
the number of particles per site in each part of the spectrum as
$n_{E_0}=N_{E_0}/N_s$, $n_{\sigma_0}=\int_{E \in \sigma_0} dE \,
\rho(E)\,\big[z^{-1} e^{\beta (E-E_0)}-1\big]^{-1}$ and
$n_{E_{+}}=N_{E_{+}}/N_s$. In the thermodynamic limit, one has
\begin{displaymath} n_{E_0}(T) = \lim_{L \to \infty}
\frac{1}{N_s} \frac{1}{z^{-1}-1},
\end{displaymath}
and $n_{E_{+}}=0$ since
\begin{eqnarray*}
n_{E_{+}}(T)&=&\lim_{L\to \infty} \frac{1}{N_s}~
\frac{1}{z^{-1} e^{-2\beta E_0}-1}<
\frac{1}{e^{-2\beta E_0}-1}= 0 \quad  \quad \forall \: T.
\end{eqnarray*}
Thus, in the thermodynamic limit, $E_{+}$  is not
macroscopically occupied at any temperature and does not play any role
in describing the thermodynamics of the system.

The last term of the right-hand side of Eq. (\ref{filling})
represents the number of bosons in the delocalized (chain-like) states.
The presence of the hidden spectrum changes the behavior of the integral
evaluated in the interval $\{-2t,2t\}$, since it reduces it to the one
describing non-interacting bosons on a linear chain with an impurity in
one of the sites. As a result, letting $z \to 1$, the integral
converges, even at finite temperatures, making possible the topology
induced spatial BEC in the center of the star graph.

\subsection{Critical temperature and condensate fraction}

If one defines $T_c$ as the critical temperature at which BEC occurs,
for any  $T<T_c$, the ground-state is macroscopically filled.
Since, at the critical temperature and in the thermodynamic limit,
$n_{E_0}(T_c)=0 $, from Eqs. (\ref{rho_catena}) and (\ref{filling}) one
has that the equation allowing to determine $T_c$ as a function of the
filling $f$ and of the hopping strength $t$ reads as
 \begin{equation}
\pi f=\int_{-2t}^{2t} \frac{dE}{\sqrt{{4t^2-E^2}} }
\frac{1}{e^{(E-E_0)/(k_{B}T_{c})}-1 }
\label{critical_T}
\end{equation}
Equation (\ref{critical_T}) can be solved numerically for any
value of $f$.  When $f \gg 1$, one may expand
the exponential in Eq. (\ref{critical_T}) to the first
order in the inverse of the critical temperature $T_c$ getting
\begin{equation}
\frac{\pi f}{k_B T_c} \approx \int_{-2t}^{2t} dE
\frac{1}{\pi \: \sqrt{4t^2-E^2}} \frac{1}{E-E_0}.
\label{critical_t2}
\end{equation}
Substituting $\cos{\theta}=E/2t$ in Eq. (\ref{critical_t2}), one has
\begin{displaymath}
\frac{2 t \pi f}{k_B T_c}=
\int_{0}^{\pi} \frac{d\theta}{\cos{\theta}-E_0/2t},
\end{displaymath}
from which
\begin{equation}
k_B T_c =E_J \sqrt{\Bigg( \frac{E_0}{2t} \Bigg)^2-1},
\label{ris_temp_crit}
\end{equation}
with $E_J$ the Josephson energy defined in 
Eq. (\ref{Jos-ij}). The result (\ref{ris_temp_crit})
holds for any graph for which $E_0<-2t$ and 
the density of states of the continuum part 
of the spectrum is given by Eq. (\ref{rho_catena}), i.e., the density
of states of a linear chain. Since, for a comb lattice $E_0=-2 \sqrt{2}
t$ \cite{burioni00}, for this graph one gets $k_B T_c =E_J$.
For a linear chain one has instead $E_0=-2t$ and thus $T_c = 0$: of
course, no condensation occurs in this case.

Upon inserting the value of the ground-state
energy (\ref{energia_ground_state}) in Eq. (\ref{critical_t2}), the
critical temperature $T_c$ is given by
\begin{equation} 
k_B T_c \approx \frac{p-2}{2\sqrt{p-1}} E_J. 
\label{t_c_f}
\end{equation}
Equation (\ref{t_c_f}) has been checked numerically and it is in
excellent agreement with the numerical solution of 
Eq. (\ref{critical_T}): for $f \gg 1$, the error is of order $1/f$.
For interwell barriers $V_0$ of order
$\sim 15 E_R \sim 2 \pi \hbar \cdot 50 kHz$ and for fillings $f \sim 200$ one
has $E_J \sim 50 nK$. According to Eq. (\ref{t_c_f}), one then
expects the formation of an observable condensate in the star
center. In Fig. \ref{fig4} we plot the critical
temperature $T_c$ given by Eq. (\ref{t_c_f}) as a function of the number
of arms.

One may use Eq. (\ref{t_c_f}) to determine also the condensate fraction
as a function of the scaled temperature $T/T_c$. In the thermodynamic
limit, the number of particles in the delocalized states is given by
\begin{equation}
N_{\sigma_{0}}(T/T_c)= \lim_{L \to \infty} N_s
\int_{-2t}^{2t} \rho(E) \frac{dE}{e^{\beta (E-E_0)}-1 }\approx N_T
\cdot \frac{T}{T_c}.
\label{n_B}
\end{equation}
In Eq. (\ref{n_B}) the exponential has been expanded to the
first order in $\beta$: this approximation
holds for $f \gg 1$ and it is in very good agreement with the numerical
evaluation of the integral (\ref{n_B}) also in a large neighborhood
below $T_c$. The critical temperature at which BEC occurs crucially
depends on the number of arms of the star (see Eq. (\ref{t_c_f})) and
thus one may adjust it by choosing a pertinent number of arms. 

From Eqs. (\ref{filling}) and (\ref{n_B}), one gets the number of
particles in the localized ground-state $N_{E_0}$: the fraction of
condensate, for $T<T_c$, is then given by \begin{equation}
\frac{N_{E_0}}{N_T} 
\approx 1 - \frac{T}{T_c}.
\label{n_0}
\end{equation}
For $f$ ranging from $10^3$ to $10^9$, the results provided by 
Eq. (\ref{n_0}) differ from those obtained by the numerical
evaluation of $N_{E_{0}}$ from Eq. (\ref{filling}) by less than $1 \%$.
Equation (\ref{n_0}) clearly shows that the condensate has dimension
$1$; cigar-shaped one-dimensional atomic Bose condensates support, in
fact, a condensate fraction given by Eq. (\ref{n_0})
\cite{ketterle96,gorlitz01}.

\subsection{Distribution of bosons along the arms of the star network}

In the following we shall determine the distribution of the bosons
over the star graph. Due to the topology induced spatial
condensation in the center of the star graph, one should expect an
inhomogeneous distribution of the bosons along the arms of the network.
The average number of bosons $N_B$ at a site $(x,y)$
depends - due to the symmetry of the graph - only on the distance
$x$ from the center of the star. At any temperature, $N_B$ is given by:
\begin{eqnarray}
N_{B}\left(x;T/T_c\right) &=& \lim_{L \to \infty}
\Big\{ N_{E_0}\left(x;T/T_c\right)\: \vert \psi_{E_0}(x)\vert^2+
 (pL+1) \int_{-2t}^{2t}dE \:\rho(E) \: \frac{1}{e^{\beta (E-E_0)}-1} \: 
\vert
\psi_{E}(x) \vert^2 \Big\}.
\label{localizzazione}
\end{eqnarray}
In Eq. (\ref{localizzazione}) $\psi_{E_0}(x)$ is the wavefunction
corresponding to the ground-state of the single-particle spectrum and
 $\psi_E(x)$ is the wavefunction associated to a delocalized state 
with energy $E$. The last term in the right-hand side of Eq.
(\ref{localizzazione}) gives then the contribution coming from the
delocalized states, which, in the thermodynamic limit, is independent
from the site index $x$ and equals the constant $\left(N_T/N_s\right)
\cdot (T/T_c)$. Using Eq. (\ref{n_0}), for $T<T_c$, one has
\begin{equation}
N_{B}\left(x;T/T_c\right) \approx \lim_{L \to \infty} N_T \left\{
\left(1- \frac{T}{T_c}\right) \frac{p-2}{2p-2}
e^{-x\log(p-1)}+
\frac{1}{p\, L+1}\cdot \frac{T}{T_c}
\right\}.
\label{rapporto1}
\end{equation}
The exponential behavior of the ground-state eigenfunction leads, for
$T<T_c$, to an increase of $N_{B}$ on the sites near the center of the
star while, when $x \gg 1$, the behavior is dominated by the last term
in the right-hand side of Eq. (\ref{rapporto1}). Thus, away from the
center, once the filling is fixed, $N_B$ depends only on the scaled
temperature $T/T_c$ and it is given by
\begin{equation}
\label{rapporto_sempl}
\frac{N_B (x;T/T_c)}{f} \approx \frac{T}{T_c}.
\label{num_bos}
\end{equation}

Topology induced spatial BEC in a system of non-interacting bosons
hopping on a star graph predicts then a rather sharp decrease of the
number of bosons at sites located away from the center. The linear
dependence exhibited by the solid line in Fig. \ref{fig5} is consistent
with the observation that, in this system, the condensate has dimension
1.

\section{Concluding Remarks}
We showed how the topology of an optical lattice confining a dilute 
atomic gas may catalyze the existence of new and unexpected coherent 
phases. For this purpose we analyzed the paradigmatic and simple example 
of bosons hopping on star graph; our analysis  allowed not only
for the computation of  the critical temperature $T_c$ for which there 
is condensation of the bosons in the center of the star, but also allowed 
us to compute the distribution of the bosons along the arms of the star 
and to show the simple dependence of the non-condensate fraction
on the reduced temperature $T/T_c$. We find $T_c \propto tf$ where $t$
is the tunneling rate and $f$ the average number of particles per site.

In this paper we analyzed the behavior of bosons hopping on a star
shaped network in the thermodynamic limit. It is comforting to observe
that numerical simulations point out to the fact that - already for $f
\sim 100$ and for a reasonable number ($L \sim 50$) of lattice sites on
each arm of the star graph - finite size effects are negligible and,
thus, the results derived in this paper are also very useful to
pertinently describe the variety of experimentally accessible systems
for which it is expected to observe the signature of a topology induced
BEC. Although we focused our attention to bosons hopping
on star shaped optical networks, the reader may easily convince
her(him)-self that our analysis could be also applied to the description
of topology induced coherent phenomena in star shaped Josephson junction
networks \cite{burioni00}. In the latter application, there is
practically no limitation on the number of junctions needed to build the
star shaped network and, thus, the results obtained in the thermodynamic
limit are expected to be very accurate.

An experimental realization of a star-shaped optical lattice
with four (or six) arms may be achieved by first creating a regular
square (or cubic) lattice using two pairs of counterpropagating laser
beams; for these configurations, the optical potential has the form
$V(x,y)=V_0 [\sin^2{(kx)}+\sin^2{(ky)}]$ (or similarly $V(x,y,z)=V_0
[\sin^2{(kx)}+\sin^2{(ky)}+\sin^2{(kz)}]$). A row and a column of the
square lattice (or three perpendicular chain of the cubic lattice) may
then be selected by superimposing two (or three) perpendicular
gaussian laser beams, obtaining by this procedure a four-
(six-) arm star optical lattice. Typical experimental values of $V_0$
for which the tight-binding approximation and the Bose-Hubbard
Hamiltonian (\ref{B-H}) are valid are $V_0 \sim 10 \, - \, 30 E_R$
($E_R=h^2/2m \lambda^2$ with $\lambda$ the wavevector of the
lattice beams, with $\lambda \sim 800 nm$). With an average number
of particle per site $f \sim 200$, from Eq. (\ref{t_c_f}) one
finds (for $V_0 \approx 15 E_E$ and $p=6$) $T_c$ of the order of $50
nK$. Below $T_c$, the macroscopic occupation of the ground-state could
be evidenced by turning off the magnetic+optical trap and observing the
gas expansion, as in the usual detection of atomic Bose-Einstein
condensates.

The experimental observation of topology induced spatial BEC on a star
shaped network may be easily achieved also using superconducting
Josephson junctions. For a superconducting network, it is sufficient to
measure the $I$-$V$ characteristic of a single arm of the Josephson
junction network (JJN) built on a star graph and of the Josephson
critical current along a given arm; if, in fact, one feeds an external
current $I_{ext}$ at the extremities of the arm, one expects to observe
no voltage unless $I_{ext}$ is larger than the smallest of the
critical currents of the junctions along the arm. Since, below $T_c$,
the Josephson critical current of the arm is given by the smallest of
the critical currents of the junctions positioned along the arm, the
measurement of the $I$-$V$ characteristic of an arm of the star graph
should provide a measurement of the critical current of the junction
which is farther from the star center. From the analysis carried out in
this paper, it is rather easy to provide an estimate of the Josephson
critical current as a function of both the temperature and the distance
from the star center above and below $T_c$. Furthermore, one may show
that the ratio of the Josephson critical currents of a Josephson
junction - located on a given arm between the sites $(x+1,y)$ and
$(x,y)$ - above and below the critical temperature $T_c$ does not depend
on $y$ and it is given by $I_c^B(x,\tau)/I_c^A(x) \approx
\sqrt{N_B(x+1;\tau)N_B(x;\tau)}/f$, where $I_c^B$ ($I_c^A$) is the
Josephson critical current below (above) $T_c$, $\tau \equiv T/T_c$ is
the scaled temperature and $N_B(x;\tau)$ is given by 
Eq. (\ref{num_bos}). Far away from the star center $(x \gg 1)$, 
one gets
\begin{equation}
\frac{I_c^B(x,\tau)}{I_c^A(x)} \approx \frac{T}{T_c} .
\label{ratio_crit}
\end{equation}
Thus, BEC in a star shaped JJN predicts a sharp decrease of the
Josephson critical current for a junction located away from the center.

The striking and intriguing similarities between superconducting
Josephson junction networks and atomic gas in suitable deep optical
lattices have been already pointed out \cite{trombettoni03}: one may
think, in fact, to realize a star shaped network also using bosonic
Josephson junctions. For this purpose, it is needed that, in each well
of the periodic potential, there is a condensate grain appearing at a
Bose-Einstein condensation temperature $T_{BEC}$ and that, when
$T_{BEC}$ is larger than all other energy scales, the atoms in the
$i$-th well of the optical lattice may be described by a macroscopic
wavefunction $\psi_i$. Thus, it becomes apparent that an optical network
may be regarded as a network of bosonic Josephson junctions. Our
analysis shows that - at a temperature $T_c<T_{BEC}$ - the topology of
the star-shaped network induces a further finite temperature transition
to a state in which the bosons spatially condense in the center of the
star.

\section*{Acknowledgements}
Discussions with M. Rasetti and A. Smerzi are gratefully acknowledged.
We acknowledge financial support by M.I.U.R. through grant No.
2001028294.

\appendix

\section{Energy spectrum on a star-shaped network}
In this Appendix we solve the eigenvalue equation (\ref{eigen_eq})
with the adjacency matrix given by Eq. (\ref{adjacency}),
describing bosons hopping on a star graph with $(pL+1)$ sites (i.e., a
star with $p$ arms having $L$ sites each).

Since non-interacting bosons on a linear chain are described by plane waves
with wave vector $k$ one has that, on each arm, an eigenstate of
Eq. (\ref{eigen_eq}), corresponding to  energy $E=-2t\cos(k)$,
may be written as:
\begin{equation}
\psi(x,y)=A_ye^{ikx}+B_ye^{-ikx}.
\label{guess_f}
\end{equation}
In Eq. (\ref{guess_f}) $y=1,\cdots,p$ is an index labeling the arm,
while  $x=0,\cdots,L$ labels the sites on
the arm. The wavefunctions described in Eq. (\ref{guess_f}) are, of
course, delocalized.

Requiring that - on each arm - $\psi(x,y)$ is a solution of the
eigenvalue equation at $x=L$ amounts to require that $A_y$,
$B_y$ and $k$ should satisfy the $p$ equations
\begin{equation} 
\label{hopping_bordo}
-t\left(A_ye^{ik(L-1)}+B_ye^{-ik(L-1)}\right)=-2t\cos(k)\left
(A_ye^{ikL}+B_ye^{-ikL}\right)\qquad y=1,\cdots,p .
\end{equation}
Furthermore, the eigenstates defined in Eq. (\ref{guess_f}) should
satisfy $(p-1)$ matching conditions in the center of the star
where the wavefunctions defined on each arm are linked; thus, one has
\begin{equation}
\label{matching}
A_y+B_y=A_{y+1}+B_{y+1} \qquad y=1,\cdots,p-1.
\end{equation}
Using Eqs.(\ref{matching}), the condition in the center gives 
one more equation
\begin{equation}
\label{hopping_centro}
-t\sum_{y=1}^{p}\left(A_ye^{ik}+B_ye^{-ik}\right)=
-2t\cos(k)\left(A_{y'}+B_{y'}\right) \qquad y'=1,\cdots,p.
\end{equation}
Equations (\ref{hopping_bordo}) and (\ref{hopping_centro}) may be
grouped in a homogeneous linear system of $2p$ equations which allows to
fix the $2p$ parameters $A_y$ and $B_y$.
Upon denoting with $M$ the ($2p \times 2p$) matrix whose elements are 
the coefficients
of the linear system given by Eqs. (\ref{hopping_bordo}) and
(\ref{hopping_centro}), requiring that
\begin{equation}
\det M=\Theta(k,L) \cdot
(1-e^{2ik(L+1)})^{p-1} \cdot \big\{(p-2)\cot(k)-p\cot{[k (L+1)]}
\big\}=0,
\label{eq_k}
\end{equation}
guarantees the uniqueness of the solution. In Eq. (\ref{eq_k}) 
$\vert \Theta(k,L) \vert=1$ for any value of $k$.

One immediately sees that the values of $k$ for which $k=n\pi/(L+1)$
(with $n=1,2,...,L$) provide a set of $L$ $(p-1)$-fold degenerate
eigenstates of Eq. (\ref{eq_k}). In addition, the solutions of the
transcendental equation
\begin{equation}
(p-2) \cot(k)-p \cot[k (L+1)]=0
\label{non_degenere}
\end{equation}
provide the values of $k$ associated to non-degenerate
eigenstates. Equation (\ref{non_degenere}) can be solved numerically and
yields a set of  $(L-1)$ non-degenerate eigenvalues corresponding to
values of $k$ which - in the thermodynamic limit - are equally spaced
and separated by a distance $\pi/(L+1)$. As a result, the set of
delocalized states is formed by $(pL-1)$ states corresponding to
energies ranging between $-2t$ and $+2t$. One can easily convince
oneself \cite{burioni01} that, in the thermodynamic limit, the
normalized density of states is given by
\begin{equation}
\rho(E)=\frac{1}{\pi \sqrt{4t^2-E^2}},
\end{equation}
as in the case of non-interacting bosons hopping on a linear chain 
(see e.g. \cite{economou83}).

Since the total number of states should equal $N_s=(pL+1)$, there are
also two localized states in the spectrum: to find them, it is
convenient to look for solutions of the eigenvalue equation
(\ref{eigen_eq}) of the form
\begin{eqnarray}
\psi_0 (x)&=& A e^{-\eta x}+ B e^{\eta x} \nonumber \\
\psi_+ (x)&=& A (-1)^x e^{-\eta x}+ B (-1)^x  e^{\eta x}
\label{loc_func}
\end{eqnarray}
corresponding, respectively, to the eigenvalues $E_{0,+}=\mp 2t \cosh
\eta$. In Eqs. (\ref{loc_func}) $A$ and $B$ are normalization constants
and $\eta \equiv 1/\xi$ is a parameter accounting for the localization
of the states. One may determine the
parameters $A$, $B$ and $\eta$ by using only the normalization condition
for $\psi_0$ and by rewriting Eq. (\ref{non_degenere}) for
$k=i\eta$. Namely, one should solve:
\begin{equation}
(p-2) \coth(\eta)-p \coth[\eta (L+1)]=0,
\label{eta_legato}
\end{equation}
together with the condition that $\sum_{x,y}\vert\psi_0(x,y)\vert^2=1$.
For $L \to \infty$, one can always set $B=0$. Equation
(\ref{eta_legato}) becomes $(p-2)\coth(\eta)-p=0 $ which is solved by
$\eta=\log(p-1)/2$ yielding $E_0=-t p/\sqrt{p-1}$ and $E_+=-E_0$.
Solving the eigenvalue equation (\ref{eigen_eq}) for $E=E_{0,+}$, one
obtains the wavefunctions of both the localized states. The normalized
eigenfunction for the ground-state is then given by:
\begin{equation}
\psi_{E_0}(x)=\sqrt{\frac{p-2}{2p-2}}e^{- x/\xi}.
\end{equation}
Since, for $L \to \infty$, the normalized density of the continuum
states is given by $\rho(E)=1/(\pi \sqrt{4t^2-E^2})$, the two localized
states do not contribute to the closure relation in the thermodynamic
limit; thus, they belong to the hidden spectrum.

\bigskip

\bigskip

\begin{figure}[ht]
\centerline{\psfig{file=fig3.eps,width=7cm}}
\vspace*{8pt}
\caption{The normalized single-particle
ground-state wavefunction for bosons hopping on a star graph as a
function of the distance $x$ from the center. The number of arms
$p$ is respectively $3$ (solid line), $6$ (dotted line), and $10$
(dashed line).}
\label{fig3}
\end{figure}

\bigskip

\begin{figure}[ht]
\centerline{\psfig{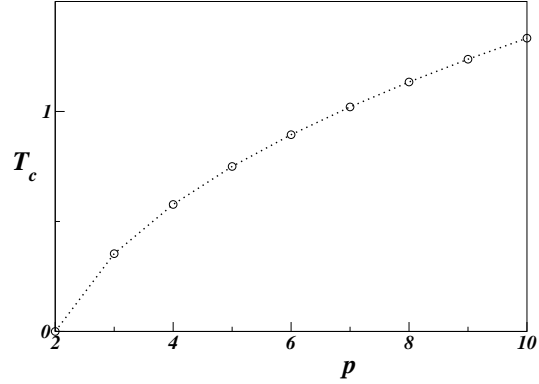}}
\vspace*{8pt}
\caption{Critical temperature $T_c$ (in units of $k_B/E_J$) 
as a function of the number of arms $p$. For the linear chain ($p=2$) 
$T_c=0$.}
\label{fig4}
\end{figure}

\bigskip

\begin{figure}[ht]
\centerline{\psfig{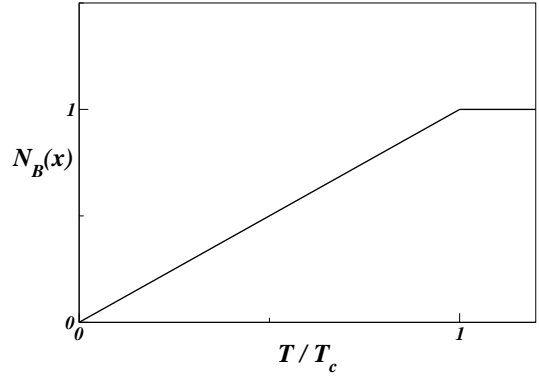}}
\vspace*{8pt}
\caption{Distribution of the number of bosons $N_B$ as a function of
$T/T_c$ computed for $x \gg 1$. $N_{B}(x)$ is in units of the
filling $f$ and is therefore equal to 1 for $T\geq T_c$.}
\label{fig5}
\end{figure}

\end{document}